\documentclass[12pt,leqno]{article}
\usepackage{graphicx}
\begin{document} 
\baselinestretch
\renewcommand{\baselinestretch}{1.5}
\noindent
\textbf{\large
Do language change rates depend on population size?}

\bigskip
\noindent
S{\o}ren Wichmann, Dietrich Stauffer, Christian Schulze, Eric W. Holman

\bigskip
\noindent
Department of Linguistics, Max Planck Institute for Evolutionary 
Anthropology, Deutscher Platz 6, D-04103 Leipzig, Germany \& Languages 
and Cultures of Indian America (TCIA), PO Box 9515, 2300 RA Leiden, The 
Netherlands. E-mail: wichmann@eva.mpg.de

\medskip
\noindent
Institute for Theoretical Physics, Cologne University, D-50923 K\"oln, Euroland.
E-mail: stauffer@thp.uni-koeln.de

\medskip
\noindent
Department of Psychology, University of California, Los Angeles, California 9009
5-1563, USA. E-mail: Holman@psych.ucla.edu

\bigskip
\noindent
\textbf{Abstract}

{\small
An earlier study (Nettle 1999b) concluded, based on computer simulations and some inferences from empirical data, that languages will change the more slowly the larger the population gets. We replicate this study using a more complete language model for simulations (the Schulze model combined with a Barab\'asi-Albert network) and a richer empirical dataset (\textit the {World Atlas of Language Structures} edited by Haspelmath et al. 2005). Our simulations show either a weak or stronger dependence of language change on population sizes depending on the parameter settings, and empirical data, like some of the simulations, show a weak dependence.}

\bigskip
\noindent
\textit{Keywords:} WALS, computer simulation, diffusion, change rate

\normalsize
\section{Introduction}
Do languages spoken by lots of people change less than those spoken by less 
people? Common sense says yes: The more people speak a language, the more 
inertia this language has, similar to the influence of mass on 
velocity changes in physics. Indeed, one of the successful computer models
for language change, the so-called Viviane model (Oliveira et al. 2006, 
Oliveira et al. 2007) of physicists, assumes this effect from the beginning.
However, human beings sometimes behave differently from inanimate atoms and
thus a direct test of this hypothesis would be desirable. 

Nettle (1999b) presents such a test for human languages.
He argues that ``spreading an innovation over a tribe of 
500 people is much easier and takes much less time than spreading one over
five million people''. His paper mainly contains a computer simulation of 
language change for just two linguistic features (the simulations can also 
be interpreted as describing the competition between two languages). 
He finds that the rate at which the majority of the
population switches between these two choices decreases to a small but 
nonzero limit if the population increases from 120 to 500. Such switching 
processes could also have been studied in the simpler Ising model of statistical
physics, where the switching rate is known (Meyer-Ortmanns and Trappenberg 1990)
to decay exponentially to zero with increasing population size. (We also found 
switching in Nettle's model at high noise level and everybody being influenced
by all speakers equally, without having to use any differences in social status,
bias, distance or age.) However, these models for only two choices cannot be
tested on the empirical language size distribution for the nearly $10^4$ 
human languages, in contrast to the later Viviane, Schulze, and 
Tuncay language competition models, to be discussed in section 2.

Our aim is to investigate whether Nettle's result may be replicated if we apply a somewhat different model than the one he described in Nettle (1999a), which was based on the Social Impact Theory of (Nowak et al. 1990). In Nettle's model, the impact of a linguistic variant is a function of the statuses of the individuals using this variant, their social distance from the learner, and their number. Our model contains parameters that are similar but not identical. Rather than assigning variable statuses to individuals we operate with a scale-free network, where the impact of a certain individual increases with a probability which is proportional to the impact that the individual already has had. Social distances correspond to distances among individuals in the network which we are using. The size of the population having a given linguistic variant indirectly affects the probability that this variant will diffuse further in one version of our model where a speaker randomly adopts variants from the entire population. The major difference between our model and Nettle's is that ours is more realistic inasmuch as it operates with many languages each of which has several features, whereas Nettle's model, depending on how one interprets it, either has one language with two competing features or two competing languages with no internal structure.

Towards the end of the paper we analyse empirical data and compare our findings with Nettle's inferences based on the empirical data which were available at his time of writing.

\section{Computer Simulation}

\subsection{Model}

Of the many computer models for language competition (Abrams and Strogatz 2003, 
Patriarca and Lepp\"annen 2004, Mira and Paredes 2005, Kosmidis et al. 2005, 
Pinasco and Romanelli 2006; Schw\"ammle 2005, see also Cangelosi and Parisi 
2002, Culicover and Nowak 2003, Pr\'evost 2003, Itoh and Ueda 2004, Wang and
Minett 2005), only the Viviane model (Oliveira 2006, 2007), the Tuncay model 
(2007) and the Schulze model (Schulze et al. 2005,
2007) gave reasonable agreement with the empirical observed distribution
of language sizes (where the size of the language is defined as the
number of people having this language as their mother tongue.) The Viviane
model assumes from the beginning that small languages change more rapidly than
large ones. The Tuncay model
does not deal with the features of a language and thus seems not suitable to 
measure language change. The most suitable model for our purposes, then, is the Schulze model, details 
of which are reviewed in the appendix.

Of the various versions of the Schulze model, we applied the one on scale-free 
social networks (Barab\'asi and Albert 1999, used for linguistics already by
Kalampokis et al. 2007) not only because it gave thus far 
the best size distribution for languages (Schulze et al. 2007). We also needed it
because we wanted to measure change rates. Normally, once a language is spoken 
by more than half of the people, it keeps that status of dominance forever in
the Schulze model. Thus, we observe a situation analogous to what Nettle called the ``threshold problem'' and described with reference to other scholars before him, such as Keller (1994), who observed that if the learner adopts the norms of his or her immediate surroundings, then the result after a few generations is always ``homogeneity if the starting point is heterogeneous and stasis if the starting point is homogeneous'' (Keller, 1994: 99, cited after Nettle 1999a: 99). Only on the scale-free networks of Barab\'asi and Albert (1999) did we observe in the Schulze model that the dominating language often changes, as has happened in Europe roughly during the course of the last two millennia, where Greek, Latin, French, and English have successively replaced the previous dominating language. In this scale-free network the most connected individuals are responsible for most, if not all, changes in the dominating language. We tested to see what happened if we disallowed modifications in the speech of the centrally connected nodes and found that the dominating language will then not change.

We made $10^3$ to $10^5$ iterations (sweeps through the network), ignored the 
first 100 of them, and counted all later changes where the role of the 
dominating language shifts from one language to another one. We counted both 
the changes in one arbitrarily selected feature (the first one) and changes 
in any of the features. If all $F = 8$ features would be independent of each
other (which they are not) and if all change rates would be small (which they
are only in some parameter regions) then the change rates for the first feature
would be eight times slower than those for the whole language (denoted by
``all'' instead of ``first'', i.e. counting the change in any of the features).
Roughly this is the case, i.e. of a pair of curves the higher one counts 
changes per language, and the lower one counts changes per feature.

\subsection{Local diffusion}

\begin{figure}[htb]
\begin{center}
\includegraphics[angle=-90,scale=0.5]{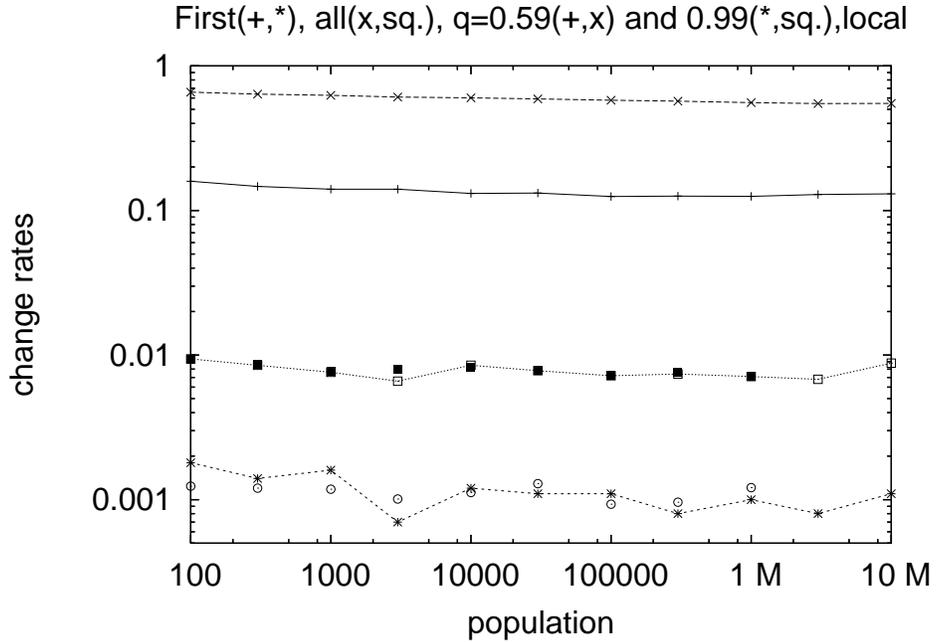}
\end{center}
\caption{Variation of language and feature change with population size, for
intermediate (higher pair of curves) and high (lower pair of curves) diffusion 
probability. The circles correspond to the stars but with $10^5$ instead of 
$10^4$ iterations.
}
\end{figure}
\begin{figure}[htb]
\begin{center}
\includegraphics[angle=-90,scale=0.5]{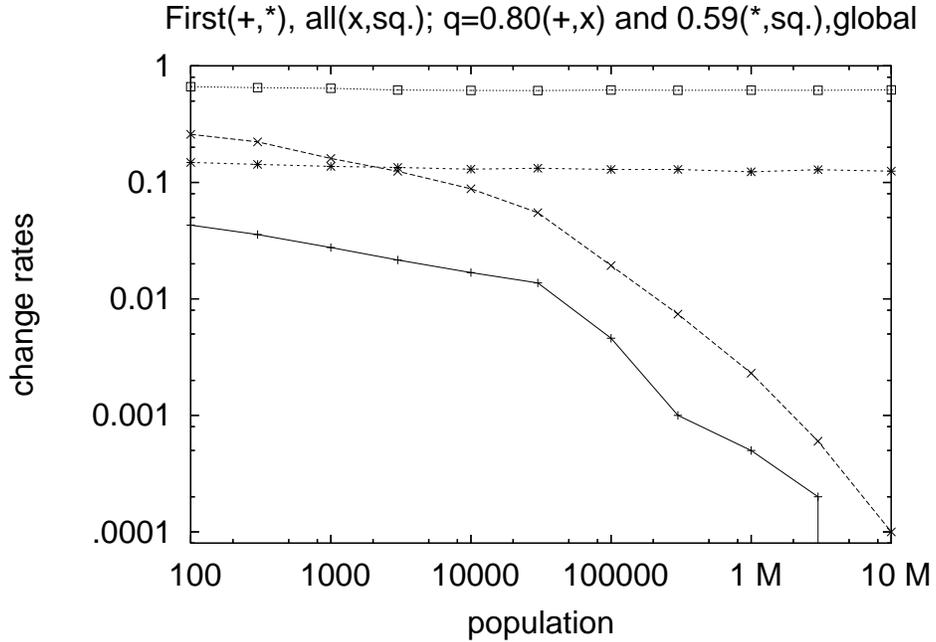}
\end{center}
\caption{As Fig.1 but now for global instead of local diffusion.
}
\end{figure}
\begin{figure}[htb]
\begin{center}
\includegraphics[angle=-90,scale=0.3]{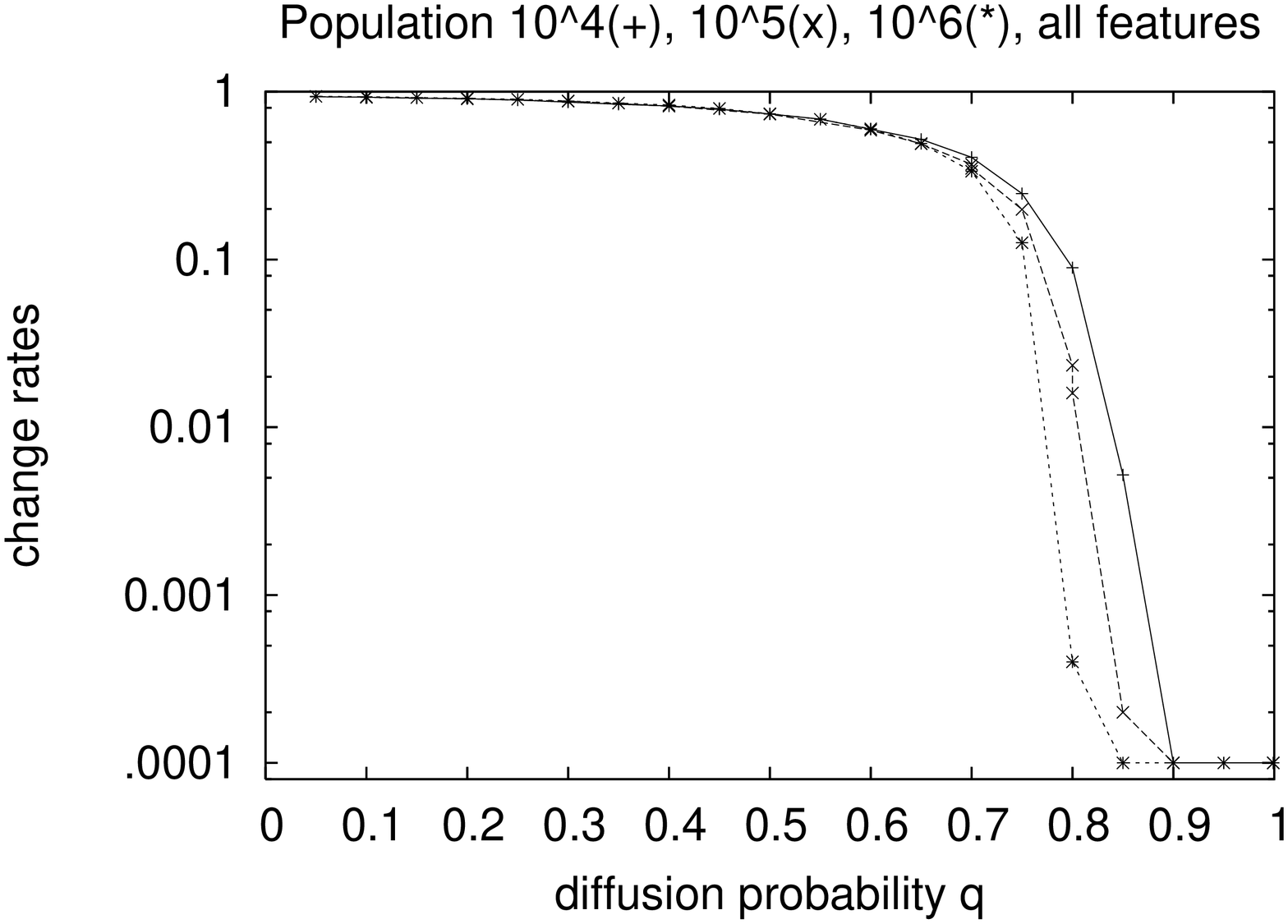}
\includegraphics[angle=-90,scale=0.3]{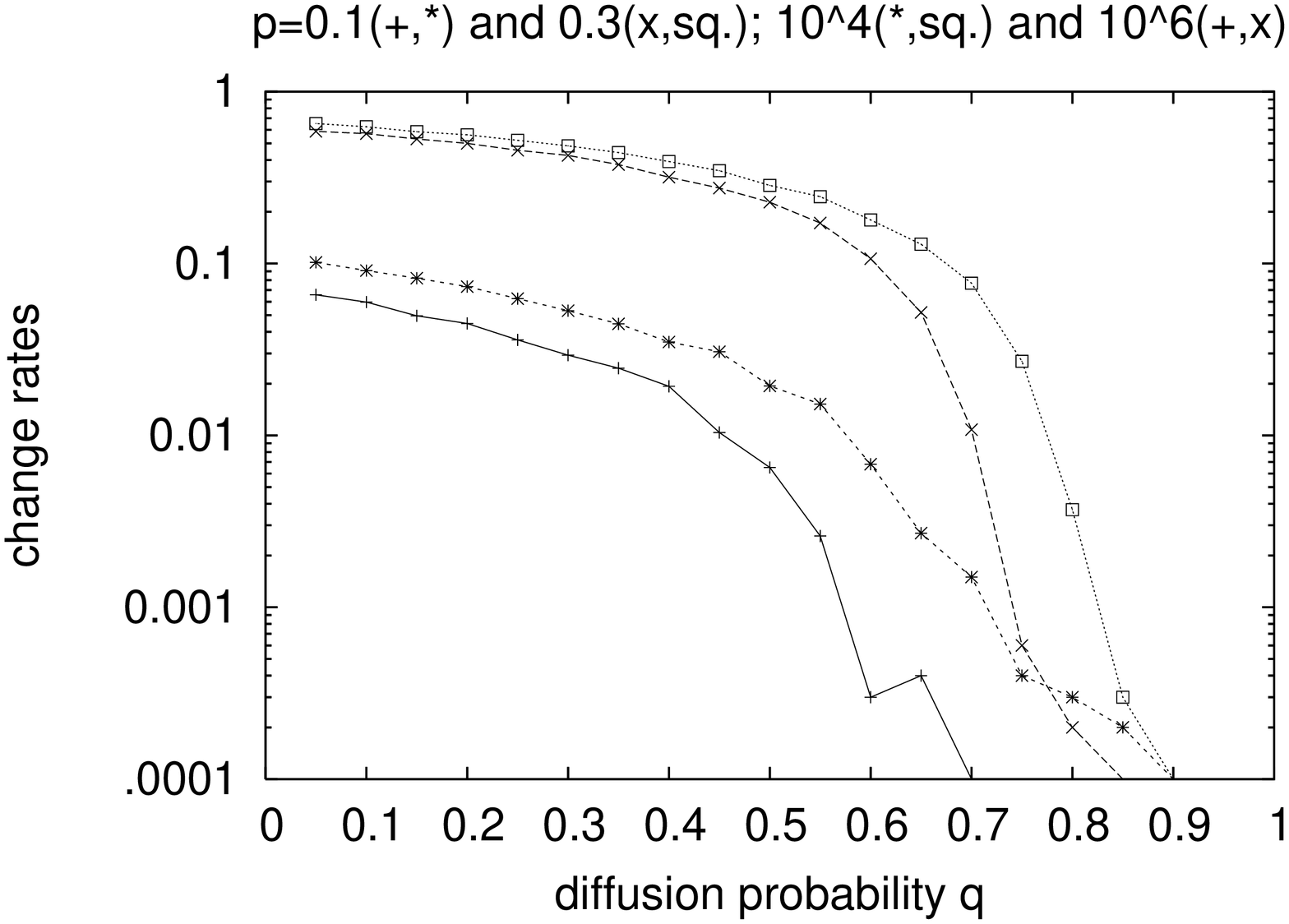}
\end{center}
\caption{Variation of language change versus diffusion probability:
$p = 0.5$, 0.3 and 0.1 from top to bottom; population $10^4 \dots 10^6$.
}
\end{figure}

To simulate the process where one speaker adopts a feature from another speaker (diffusion, transfer) we distinguish between a situation where the donor can sit on a neighbour node of the network (``local version'') or on any randomly selected node (``global version''). We will start with the local version and then present the global one. 

Fig.1 shows for the local version that neither at intermediate nor at high
diffusion probabilities $q$ is there a strong variation of change rates 
with population sizes (= number of network nodes) varying over five decades 
from 100 to 10 million. However, the stronger coupling between languages
at high $q$, compared to low $q$, makes language change more rare. So, these
local simulations give a clear answer: Population size has little influence.

\subsection{Global diffusion}

For global diffusion the situation is quite different. Now for 
intermediate $q$ again no clear influence of populations size can be seen 
in Fig. 2, but for larger $q$ the rate of change is diminished drastically
with increasing population size. Fig.3 confirms this picture over the whole
range of $q$: Only for large $q$ when the change rates become small does the 
population size have a strong influence on them. (The analogous figure with 
local instead of global diffusion has overlapping data, cf. Schulze et al. 2007.) The size 
effects in the bottom part are in the same direction but stronger than in 
the top part. (Migration via exchange of nearest neighbours had little
influence.)

In the results for both the local and the global situation just reported the complete language shift remains local. If we change from local to global interactions in the shift from one language to the other, then the dominating language always retains 
its dominating position and no change happens.

One can argue that for small populations global and local diffusion are more 
similar than for large populations and thus the distinction is less important.
Only for large populations does global diffusion give change rates different 
from their high values for local diffusion.

\section{Empirical data}

\begin{figure}[htb]
\begin{center}
\includegraphics[angle=-90,scale=0.5]{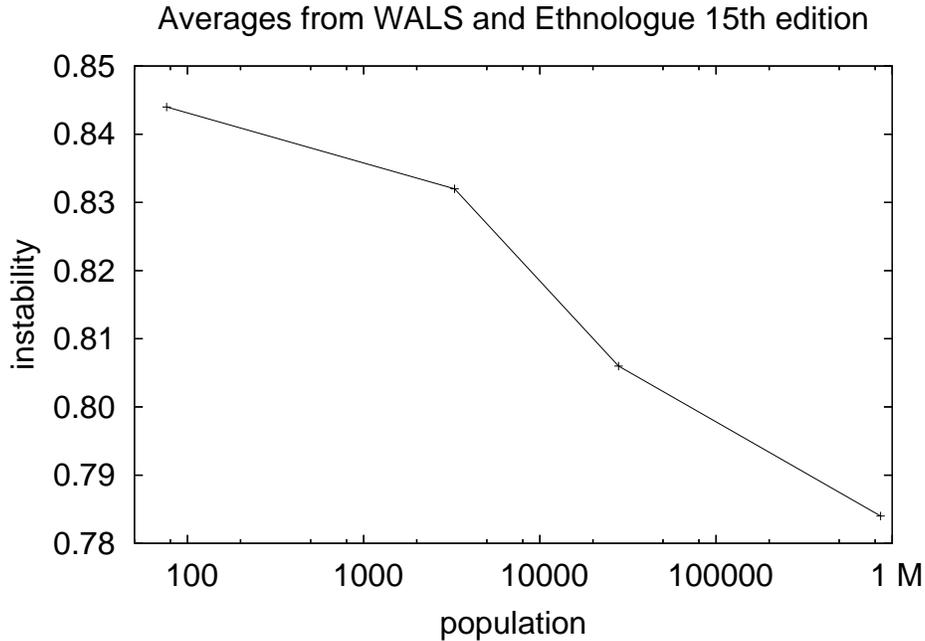}
\end{center}
\caption{Average instability (= 1 minus stability) versus average 
population size for real language families.
}
\end{figure}

\begin{figure}[htb]
\begin{center}
\includegraphics[angle=-90,scale=0.5]{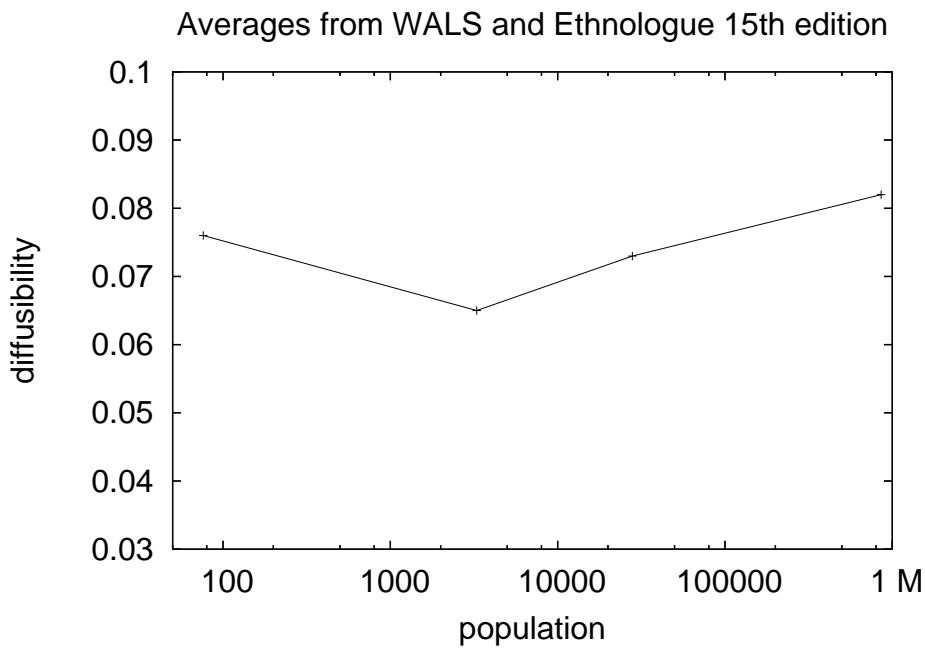}
\end{center}
\caption{Diffusibility (see text) versus average population size for real 
language families. 
}
\end{figure}

We used the \textit{World Atlas of Language Structures} (WALS, Haspelmath et al. 2005), which maps structural features across a number of the world's languages,
together with the \textit{Ethnologue} language statistics (Gordon 2005), to estimate the instabilities and thus indirectly the change rates of real languages.

To study rate of change as a function of number of speakers, the number of speakers of each language was obtained from \textit{Ethnologue}.  Extinct languages and those with unknown numbers of speakers were omitted from the sample.  The remaining languages were divided into four groups of approximately equal size: 1 to 999 speakers (423 languages), 1000 to 9999 speakers (513 languages), 10,000 to 99,999 speakers (549 languages), and 100,000 or more speakers (655 languages).

One way to infer rate of change is from the instability of linguistic features by assuming that the more unstable the features are, the faster the rate of change is.  Accordingly, the instability of each of the 134 nonredundant WALS features was estimated in each of the four groups of languages.  The measure of instability is adapted from Wichmann and Holman (n.d.). (See Holman et al. 2007 for a summary).  Within each language family, we look at all the pairs of languages located within 5000 km of each other for which a given feature is attested in both languages, and find the proportion $R$ of such pairs for which the feature has the same value.  We do the same for languages in different families, getting the proportion $U$.  Then $(1-R)/(1-U)$ is the instability. Instability as defined here is equal to one minus stability as defined in Wichmann and Holman, except that related languages are here considered to be those in the same family rather than those in the same genus, in order to maximize the number of related languages in each group.  Thus the values of instability are higher than those inferred from Wichmann and Holman, and the two cannot be compared in absolute terms.  Fig. 4 plots the mean stability of the features as a function of the geometric mean number of speakers in the group.

The figure shows a slight decrease in instability with number of speakers.  To determine whether the decrease is statistically significant, the Spearman rank correlation between instability and number of speakers was calculated separately for each of the 47 independent features identified by Holman (n.d.).  The mean correlation is --0.18, with a standard deviation of 0.60; the mean is significantly negative, $t(46) = 2.05, \; p < 0.05$.  However, the change is much smaller than in Figs. 2 and 3, and this reality is closer to Fig. 1 based on local diffusion, where also in the population range between 100 and $10^6$ a slight decrease of change rates with increasing population was found. 


We have also investigated whether diffusibility is dependent on population sizes. Diffusibility is logically independent of stability, since a particular feature may remain stable for a long time after it has diffused. Support for the independence of the two phenomena is provided in Wichmann and Holman (n.d.), where it is demonstrated that individual features have comparable stabilities across languages, while the diffusibility of a given feature may vary from area to area. 

The diffusibility of each feature was estimated in each of the four groups of languages, again following the procedure described in Wichmann and Holman (n.d.).  Among all the pairs of languages in different families located within 5000 km of each other, we look at the pairs for which a given feature is attested in both languages, and find the proportion $R$ of such pairs for which the feature has the same value.  We do the same for pairs of languages in different families located more than 5000 km each other, getting the proportion $U$.  Then $R/(1-U)$ is the diffusibility.  Fig. 5 plots the mean diffusibility of the features as a function of the geometric mean number of speakers in the group. 

The figure shows little change in diffusibility with number of speakers.  The rank correlation between diffusibility and number of speakers has a mean of 0.08 for the 47 independent features, with a standard deviation of 0.66, indicating no significant correlation, $t(46) = 0.84$.

We believe that these empirical findings based on systematic analysis of a large dataset are more solid than the more indirect inferences of Nettle (1999b) and Wichmann (in press) concerning change rates and population sizes. Nevertheless it is interesting that these inferences point in the same direction as the findings based on WALS. Nettle (1999b: 131) observed differences in sizes of languages families in the two hemispheres and explained this by differences in change rates which would in turn be explained by differences in population sizes:

{\small
If languages are changing fast internally, then
after they split, identifiable relationship will be quickly erased from their descendants,
and so, after a given time period, there will appear to be many, small language
families. If the languages are changing very slowly, then identifiable relationship
will persist for longer, and so the reconstructable language families will be much
broader. In short, a slow rate of change predicts the Old World situation, with few
families each of which has many members, whilst a fast rate of change predicts the
New World situation, with many families each of which contains few languages.}

A modified version of this argument was presented in Wichmann (in press), where not only the number of languages in different families ($n$) was taken into account, but also the diversity within families ($d$). For diversity measures glottochronological dates were used. The ratio $n/d$ was labelled the `density' ($D$). A correlation was found between small values of $D$ and (present or erstwhile) hunter-gatherer societies and high values of $D$ and sedentary societies whose subsistence has been based on agriculture or fiver fishing. Thus, it was argued, hunter-gatherers tend to live in smaller groups and their languages tend to change faster than sedentary peoples.

We now briefly summarize this section's findings. Population size has no systematic effect on diffusibility. The degree to which languages undergo contact-induced change is probably ultimately dependent on particular histories of interaction among speakers. Internal language change, however, is more constant across language. Nevertheless, our findings show that there is a small but significant effect of language size on the rate of change such that large populations lead to somewhat slower rates of change. More circumstantial empirical evidence discussed in Nettle (1999b) and Wichmann (in press) points in the same direction.

\section{Conclusion}

Common sense or analogy with physics and with biology (Oliveira et al. 2006, 
Sutherland 2003) do not always work: Larger ``masses'' are not necessarily 
less mobile, if we identify mass with population size and mobility
with language change. We found that only for global as opposed to local 
diffusion, and for large as opposed to small diffusion 
probabilities, the rate of language change goes down drastically if the 
population size increases from 100 to ten million.

We found that only diffusion at the global 
level may have a size effect. Given a situation where (a) individuals may adopt linguistic features from individuals anywhere in the speech community, (b) certain individuals become more connected than others, and (c) diffusion is high, an increased population size will give a lowered change rate. We can then predict that languages like English 
or Mandarin Chinese will change more slowly than smaller languages spoken by populations in relative isolation from one another, as we might imagine the situation to have been for some traditional societies. But between these extremes there is a vast gray area of intermediate situations where our simulations hold little predictive power because our parameter values for diffusion, population sizes etc. are abstract and cannot be translated into precise numbers.

Here the empirical data aid us. They indicate that the conditions (a-c) mentioned above have never been present to such an extent that, over the course of recent millenia, smaller languages have changed much faster than larger ones. Nevertheless, a small but significant effect of population sizes on language change has been observed, supporting the claims of Nettle (1999b), and this should be taken into account when attempts are made to date prehistoric linguistic events.

\section{Appendix}

A language (more precisely, a grammar) in the Schulze model is defined by $F$ 
features each of which has one of $Q$ values $1, 2,  \dots Q$. (We follow Holman
et al. 2007 and use $F=8, \; Q=5$.) It evolves in discrete time steps $t=1,2,
\dots$; all individuals are updated once in each iteration.

All versions of the Schulze model are based on change, diffusion, and shift,
using three probabilities $p,q,r$ at each iteration:
\begin{itemize}
\item {\it Change:}  Each feature with probability $p(1-q)$
is randomly changed to a new value between 1 and $Q$. 
\item {\it Diffusion:} Each feature with probability $pq$ is 
replaced by the corresponding feature from a randomly selected neighbour.
\item {\it Shift:} Each individual with probability $r(1-x)^2$
gives up its language and instead shifts to the language of a randomly 
selected neighbour.
\end{itemize}
Here, $x$ for the shift is the fraction of people in the whole population
speaking the language of the individual considering a shift. Linguistically
these three types of modification may correspond to the
analog of biological mutations, to the transfer of linguistic features (loanwords or structural features) from one language
into another, and to the adoption of a new language, for instance by immigrants.

The simulation first determines, with probability $p$, whether to modify the
language, and then does it with probability $q$ by learning from a neighbour
and with probability $1-q$ by random change. In our ``local'' version this
neighbour is a nearest neighbour of the considered site, in our ``global''
version it can be any member of the population. The shift is a conscious 
decision to give up the own language in favour of a more widespread one.

Usually the individuals sit on sites of a square lattice, but for the present
paper they sit on the nodes of a ``scale-free'' Barab\'asi-Albert network. On
these lattices different nodes have different numbers $k$ of neighbours, with
a probability proportional to $1/k^3$. These networks are constructed as
follows: We start with $m$ nodes having each other as neighbours. Then new 
members join the network one after the other. They select as their own 
neighbours (more precisely: teachers) $m$ already existing network members,
with a probability proportional 
to the number of cases where these teachers were selected 
before by earlier members of the network. Thus popular nodes become even
more popular, and unpopular nodes have little chance of becoming selected later.
We used these networks instead of square lattices since on lattices the
dominating language no longer changes once the majority of people speak it.
Mostly we take $m=3$. Our networks are directed, that means if a later node A 
has selected node B as a teacher ( = neighbour), then B has not selected node A
as a teacher. 

We start with everybody selecting randomly one of the $Q^F = 390625$ possible languages.
At each iteration we determine $q_{\max}$ as the most-often spoken value of 
the first feature, and $L_{\max}$ as the most-often spoken language. Then we 
check how often $q_{\max}$ changes and denote this probability by ``first''
in some figures. Analogously we count how often $L_{\max}$ changes and mark
these probabilities by ``all'' since all features together determine a language.
All changes during the first 100 iterations were ignored. Thus we find the 
rates at which $q_{\max}$ and $L_{\max}$ change in the stationary regime,
while the input parameter $p$ gives the rate at which any feature is modified.

(We also made some tests with $F=1, Q=2$ close to Nettle's model, but then
the results were less clear than for our $F=8, Q=5$.)

\bigskip
{\bf \Large References}

\medskip
\noindent
Abrams, D. and S. H. Strogatz (2003). Modelling the dynamics of language death. 
\textit{Nature} 424: 900.

\medskip
\noindent
Barab\'asi, L.A. and R. Albert (1999). Emergence of scaling in random networks.
\textit{Science} 286, 509-512.

\medskip
\noindent
Cangelosi, A. and D. Parisi (eds.) (2002). \textit{Simulating the Evolution of Language}. Berlin: Springer-Verlag.

\medskip
\noindent
Culicover, P. and A. Nowak (2003). \textit{Dynamical Grammar}, Oxford University 
Press, Oxford

\medskip
\noindent
Gordon, R. G., Jr. (ed.). (2005). \textit{Ethnologue: Languages of the World}, Fifteenth edition. Dallas, Tex.: SIL International. Online version: \\
http://www.ethnologue.com/.

\medskip
\noindent
Haspelmath, M., M. Dryer, D. Gil, and B. Comrie (eds.) (2005). \textit{The World Atlas of Language Structures}. Oxford: Oxford University Press.

\medskip
\noindent
Holman, E. W. (no date). Approximately independent typological features
of languages. Submitted for publication.

\medskip
\noindent
Holman, E. W., C. Schulze, D. Stauffer, and S. Wichmann (2007). On the relation 
between structural diversity and geographical distance among languages:
observations and computer simulations. Conditionally accepted by \textit{Linguistic Typology}. Preprint available at http://email.eva.mpg.de/$\sim$wich\-mann/HolmanRevisedSubmit.pdf

\medskip
\noindent
Itoh, Y. and S. Ueda (2004). The Ising model for changes in word ordering rules
in natural languages. \textit{Physica D} 198: 333-339.

\medskip
\noindent
Kalampokis A., K. Kosmidis, and P. Argyrakis (2007). Evolution of vocabulary on
scale-free and random networks. \textit{Physica A} 379: 665-671.

\medskip
\noindent
Keller, R. (1994). \textit{On Language Change: The Invisible Hand in Language}. London: Routledge.

\medskip
\noindent
Kosmidis, K., J.M. Halley, and P. Argyrakis (2005). Language evolution and population dynamics in a system of two interacting species. \textit{Physica A} 353: 595-612. 

\medskip
\noindent
Meyer-Ortmanns, H. and T. Trappenberg (1990). Surface tension from 
finite-volume vacuum tunneling in the 3D Ising model. \textit{Journal of Statistical
Physics} 58: 185-198.

\medskip
\noindent
Mira, J. and A. Paredes (2005). Interlinguistic simulation and language death dynamics. \textit{Europhysics Letters} 69.6: 1031-1034.

\medskip
\noindent
Nettle, D. (1999a). Using social impact theory to simulate language change. \textit{Lingua} 108: 95-117.
	
\medskip
\noindent
Nettle, D. (1999b). Is the rate of linguistic change constant? \textit{Lingua} 108: 119-136.

\medskip
\noindent
Nowak, A., J. Szamrej, and B. Latane (1990). From private attitude to public opinion: A dynamical theory of social impact. \textit{Psychological Review} 97, 362-376.

\medskip
\noindent
Oliveira, V. M. de, M. A. F. Gomes, and I. R. Tsang (2006). Theoretical model for the evolution of linguistic diversity. \textit{Physica A} 361: 361-370.

\medskip
\noindent
Oliveira, P. M. C. de, D. Stauffer, F. S. W. Lima, A. O. Sousa, C. Schulze,
and S. Moss de Oliveira (2007). Bit-strings and other modifications of Viviane 
model for language competition. \textit{Physica A} 376, 609-616.
 
\medskip
\noindent
Patriarca, M., T. Lepp\"annen (2004). Modeling language competition. \textit{Physica A} 
338: 296-299.

\medskip
\noindent
Pinasco, J. P., L. Romanelli (2006). Coexistence of languages is possible. \textit{Physica A} 361: 355-360.

\medskip
\noindent
Pr\'evost, N. (2003). textit{The Physics of Language: Towards a Phase Transition of Language Change.} PhD dissertation, Simon Fraser University.

\medskip
\noindent
Schulze, C. and D. Stauffer (2005). Monte Carlo simulation of the rise and fall of languages. \textit{International Journal of Modern Physics C} 16: 781-787.

\medskip
\noindent 
Schulze, C., D. Stauffer, and S. Wichmann (2007).  Birth, survival and death of 
languages by Monte Carlo simulation. Conditionally accepted by \textit{Communications in Computational Physics},.

\medskip
\noindent
Schw\"ammle, V. (2005). Simulation for competition of languages with an ageing 
sexual population. \textit{International Journal of Modern Physics C} 16: 1519-1526.

\medskip
\noindent
Sutherland, W. J. (2003). Parallel extinction risk and global distribution of 
languages and species. \textit{Nature} 423: 276-279.

\medskip
\noindent
Tuncay, \c{C}. (2007). Physics of randomness and regularities for cities, 
languages, and their lifetimes and family trees. \textit{International Journal 
of Modern Physics C} 18, in press. 

\medskip
\noindent
Wang, W. S. Y., Minett J. W. (2005). The invasion of language: emergence, change and death. \textit{Trends in Ecology and Evolution} 20.5: 263-296.

\medskip
\noindent
Wichmann, S. In press. Neolithic linguistics. In: Barjamovic, G., I. Elmerot, A. Hyllested, B. Nielsen, and B. Okholm Skaarup (eds.), \textit{Language and Prehistory of the Indo-European peoples---A Cross-Disciplinary Perspective}. Budapest: Archaeolingua.

\medskip
\noindent
Wichmann, S. and E.W. Holman (no date). Assessing temporal stability for linguistic typological features. Submitted. Preprint available at http://email.eva.mpg.de/$\sim$wichmann/WichmannHolmanIniSubmit.pdf.

\end{document}